\documentclass[preprint,12pt]{elsarticle}

\usepackage{epsfig}

\usepackage{amssymb}

\usepackage{color}
\newcommand{\MARKI}[1]{#1}

\journal{Acta Astronautica}

\begin{document}

\begin{frontmatter}



\title{Photonic spin control for solar wind electric sail}


\author[FMI]{Pekka Janhunen\corref{cor1}}
\ead{pekka.janhunen@fmi.fi}
\ead[url]{http://www.electric-sailing.fi}

\address[FMI]{Finnish Meteorological Institute, Helsinki, Finland}
\cortext[cor1]{Corresponding author}

\begin{abstract}
The electric solar wind sail (E-sail) is a novel, efficient propellantless
propulsion concept which utilises the natural solar wind for spacecraft
propulsion with the help of long centrifugally stretched charged
tethers. The E-sail requires auxiliary propulsion applied to the tips of the
main tethers for creating the initial angular momentum and possibly for
modifying the spinrate later during flight to counteract the orbital Coriolis
effect and possibly for mission specific reasons. We introduce the possibility
of implementing the required auxiliary propulsion by small photonic blades
(small radiation pressure solar sails). The blades would be stretched
centrifugally. We look into two concepts, one with and one without
auxiliary tethers. The use of \MARKI{small} photonic \MARKI{sails} has the benefit of
providing sufficient spin modification capability for any E-sail
mission while keeping the technology fully propellantless.
We conclude that \MARKI{small} photonic \MARKI{sails} appear to be a feasible and attractive
solution to E-sail spinrate control.
\end{abstract}


\end{frontmatter}


\begin{figure}
\centerline{\includegraphics[width=0.45\columnwidth]{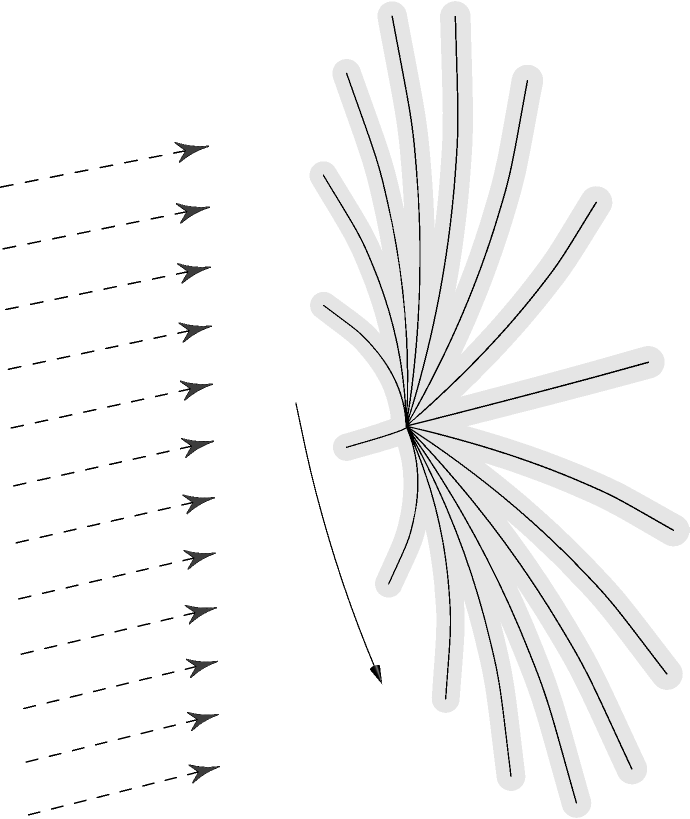}}
\caption{
Spinning E-sail in the solar wind. The solar wind force bends the
charged main tethers. The tethers are surrounded by the electron
sheaths which are shown schematically by shading.
}
\label{fig:Esailbare3D}
\end{figure}

\section{Introduction}

The solar wind electric sail (E-sail) is a newly discovered way of
propelling an interplanetary spacecraft by employing the thrust
produced by the natural solar wind plasma stream
\cite{paper1,Esailpatent}. The solar wind dynamic pressure is tapped
by long, thin, centrifugally stretched and positively charged tethers
(Figure \ref{fig:Esailbare3D}). According to numerical estimations,
the E-sail could produce $\sim$500 nN/m thrust per unit length
\cite{paper6}. It would seem possible to build e.g.~a system with 2000
km total tether length (for example with 80 tethers 25 km long each)
whose total mass is $\sim 100$ kg and which produces $\sim$1 N of
thrust at 1 AU \cite{RSIpaper} where AU is the astronomical unit.  The
thrust scales as $1/r$ where $r$ is the solar distance
\cite{paper6}. The predicted performance (1 N thrust at 1 AU, 100-200
kg mass) is high enough that it would enable a large class of
previously unattainable missions in the solar system such as sending a
~200 kg probe at more than 50 km/s speed out of the solar system to
make in situ measurements of interstellar space beyond the heliopause
\cite{RSIpaper}.

When compared with the solar photon sail, the solar wind used by the
E-sail has $\sim$5000 times smaller dynamic pressure than the
radiation pressure ($\sim$2 nPa versus 9 $\mu$Pa at 1 AU). However,
the E-sail has the crucial benefit that it uses a ``virtual'' sail
area made of the static electric field which can be $\sim$6 orders
magnitude wider ($\sim$100 m) than the physical width of the tether
wires ($\sim$100 $\mu$m). Therefore, the efficiency (thrust versus
mass) of the electric sail can be at least an order of magnitude
better than that of a solar sail using comparable materials. As
further benefits, the E-sail can control the thrust vector magnitude
and direction independently of each other (the magnitude control is
unlimited between zero and some maximum and the direction control
capability is $\sim\pm$30$^{\rm o}$).

All E-sail designs require a method of guiding the tethers so that
they do not collide with each other despite solar wind variations.
Different approaches can be used. One of the methods uses auxiliary
tethers which connect the tether tips \cite{RSIpaper}. 

For starting and possibly later modifying the spin, small thrusters
are needed in the Remote Units that are placed at the tips of the main
tethers.  In this paper we present and analyse the possibility that
these thrusters are implented as small photon sails. We first review
the E-sail with auxiliary tethers in general terms, then introduce the
photonic thrusters at the tips. After that we look into the
possibility of leaving out the auxiliary tethers. The paper ends with
a discussion and outlook of the various technical possibilities to
implement an electric solar wind sail.

\section{E-sail with auxiliary tethers and Remote Units}

The E-sail consists of the main spacecraft from which a number of
centrifugally stretched main tethers extend outward
(Fig.~\ref{fig:Esailbare3D}). The large tether rig spins slowly so
that the centrifugal force keeps the main tethers taut while the solar
wind pushes on them. In a full-scale mission there could be 100
tethers each of which is 20 km long.

\begin{figure}
\centerline{\includegraphics[width=0.5\columnwidth]{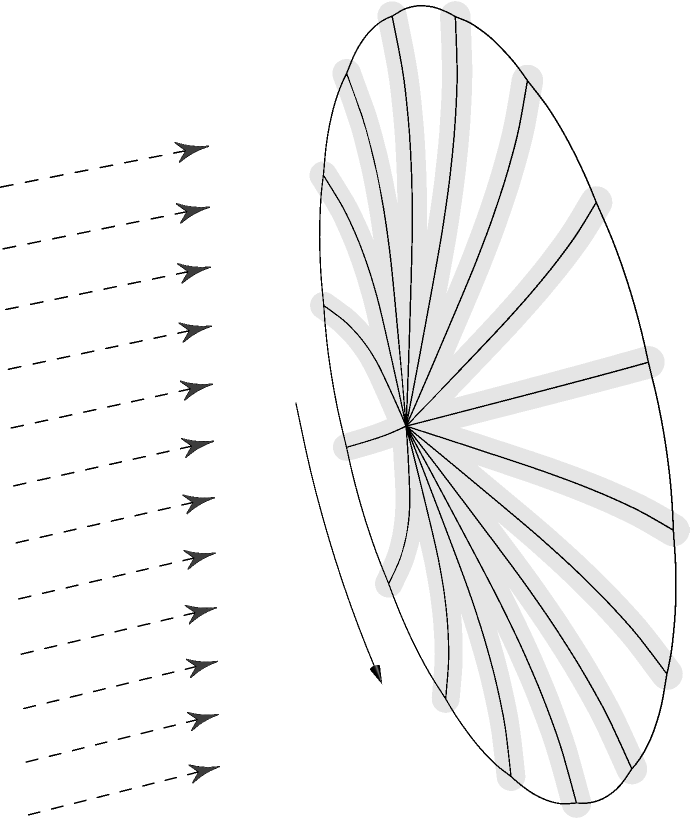}}
\caption{
E-sail with auxiliary tethers preventing the main tethers from
colliding each other despite solar wind variations.
}
\label{fig:Esail3D}
\end{figure}

The solar wind varies in time and the E-sail must usually be inclined
with respect to the flow to give the desired thrust vector
orientation. For this reason the different tethers experience a
slightly different solar wind thrust history. This would cause the
tethers to eventually spin at slightly different rates, leading them
to collide with each other. Any E-sail design must resolve this
problem in some way or another. Our baseline approach to solve it is
to connect the tips of the main tethers together by non-conducting
auxiliary tethers (Fig.~\ref{fig:Esail3D}). The tip of the main tether
must then contain a ``Remote Unit'', a small autonomous spacecraft
which hosts the reel or reels from which the auxiliary tethers are
deployed. The Remote Unit must also contain the propulsion system
which produces the angular momentum of the tether rig during its
initial deployment. The delta-v requirement for the Remote Unit
propulsion system is typically few tens of m/s. Two alternative
propulsion systems are under investigation and prototyping in the
ESAIL FP7 project: a gas thruster and an ionic liquid field effect electric propulsion (FEEP)
thruster \cite{MarcuccioEtAl2009}.

Besides producing the initial spin, a need may arise later during
propulsive E-sail flight to alter the spinrate. Particularly, if the
mission goes around the sun while spiralling inward or
outward by keeping the E-sail inclined with respect to the radial
solar wind flow, the Coriolis force due to orbital motion results in a net acceleration or
deceleration of the E-sail's spin for outward or inward spiralling orbit,
respectively \cite{paper14}. This secular change of the spinrate
$\omega$ is given by
\MARKI{
\begin{equation}
{d\omega(t)\over dt} = \omega \Omega \tan\alpha,
\label{eq:coriolis1}
\end{equation}
\begin{equation}
\omega(t) = \omega(0) e^{\Omega t \tan\alpha}
\label{eq:coriolis2}
\end{equation}
}
where $\Omega$ is the angular frequency due to orbital motion ($\Omega
= 2\pi/\tau$ where $\tau$ is the orbital period) and $\alpha$ is the
inclination angle of the sail, taken positive (negative) for
orientation causing outward (inward) spiralling orbit. There are some
indications that one could to some extent or possibly even fully
compensate for the secular spinrate change by modulating the tether
voltages in a certain way to utilise the natural small variations in
the solar wind direction \cite{paper14}. Nevertheless, it is prudent
to investigate such Remote Unit propulsion systems which allow one to
manage the spinrate during flight at will. A cold gas thruster does
not have enough delta-v capability for this task, although it is
sufficient for producing the initial spin. The ionic liquid FEEP
thruster has the potential to accomplish it, however, at least for a
substantial class of potential E-sail missions. The Remote Unit
propulsion systems are not single failure points because they can back
up each other.

\section{Remote Units with photonic blades}

\begin{figure}
\includegraphics[width=0.75\columnwidth]{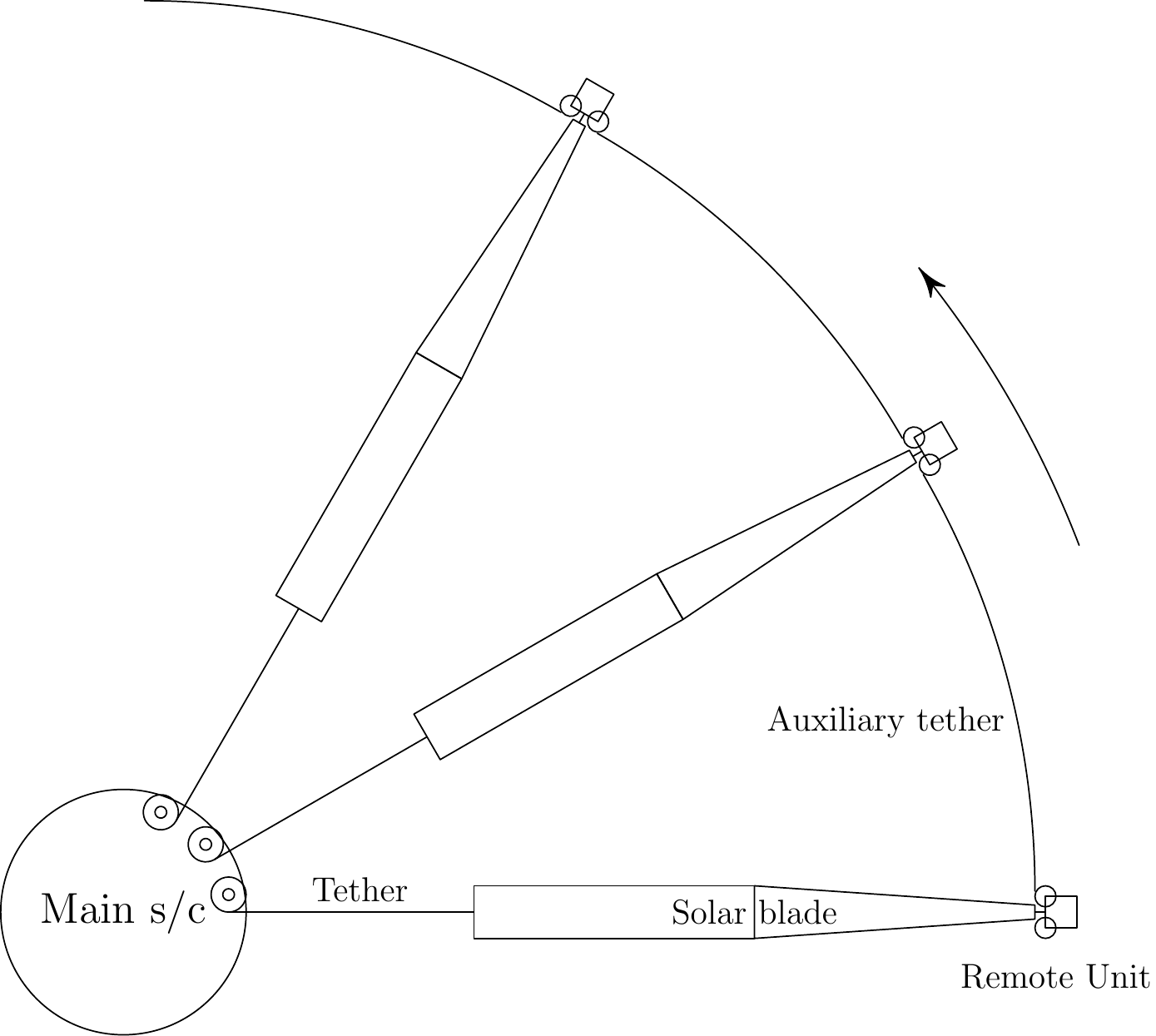}
\caption{
Schematic description of E-sail with photonic blade propulsion systems
for spinrate management. Between each main tether and the Remote Unit there is a
centrifugally tightened photonic blade whose solar angle can be controlled
mechanically from the Remote Unit.
}
\label{fig:Esail2D}
\end{figure}

\begin{figure}
\includegraphics[width=0.75\columnwidth]{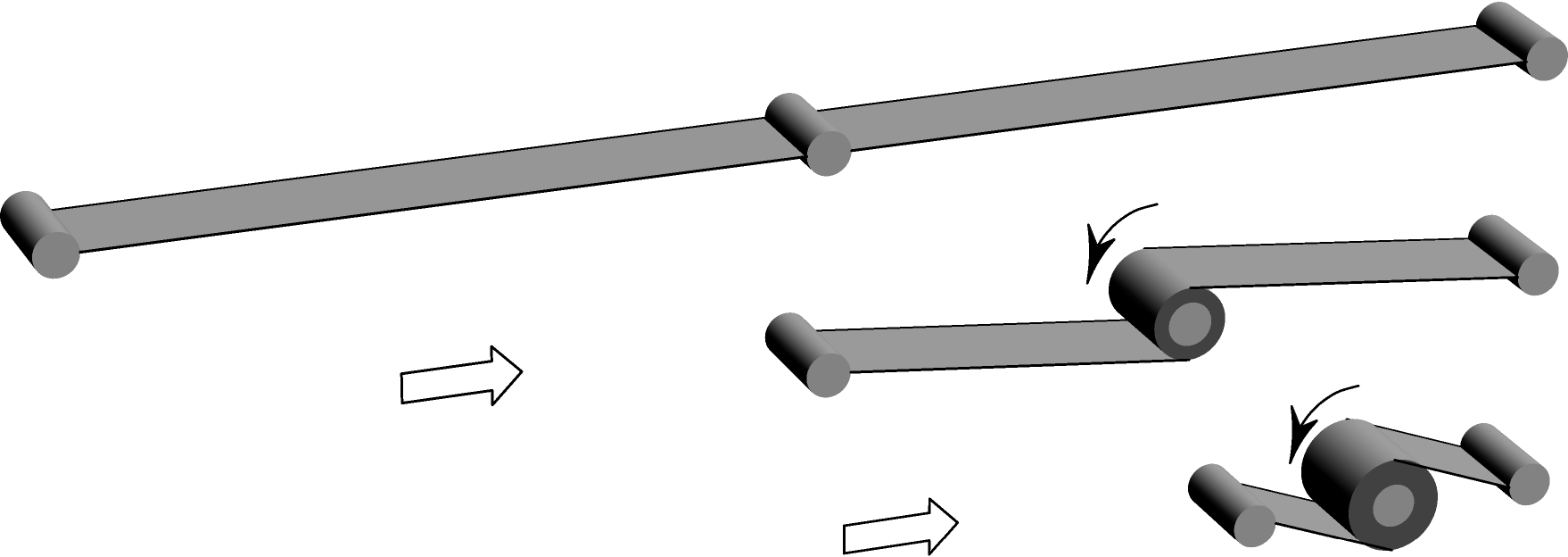}
\caption{
For the stowed configuration the solar blades are folded around a
rigid bar. Other two rigid bars are installed at the ends to keep the
blade in shape. For simplicity the narrowing of the blade at the
Remote Unit end (Fig.~\ref{fig:Esail2D}) is not shown.
}
\label{fig:rolls3D}
\end{figure}

While other solutions exist as explained above, using small photon
sails for managing the E-sail spinrate would be an attractive option
because photon sails are propellantless (like the E-sail itself) and
because at the required performance level they are potentially simple
and inexpensive. Figure \ref{fig:Esail2D} shows a concept where
flexible surface solar blades are deployed between the main tether and
the Remote Unit. In the stowed configuration the blade is rolled
around a central bar or stick, with additional rigid members keeping
it in rectangular shape after deployment (Fig.~\ref{fig:rolls3D}). The
blade is stored with the Remote Unit and it is deployed by the
centrifugal force when its holding mechanism is released. To obtain
spinrate changing torque from the blades they are tilted with respect
to sun. The tilting is actuated by the Remote Unit. The Remote Unit
will not counterrotate by the blade tilting because the auxiliary
tethers keep its attitude constant.

Depending on the size of the main spacecraft and the number of
tethers, the blade deployment could occur when some tens of metres of
the main tether has been deployed so that the blades can deploy
themselves without touching each other. The piece of main tether which
is out when blade deployment is performed must be strenthened to
withstand the mechanical shock resulting from the blade
tightening. For example, it could be made of polyimide tape because
that part of the main tether does not necessarily have to be
conducting. On the other hand, depending on the radius and moment of
inertia of the main spacecraft, the number of tethers and the length
of the blades, it might be feasible to perform blade deployment
directly from the main spacecraft. In this case the release mechanism
of the blades can be the same mechanism which releases the Remote
Units from the main spacecraft and no specially strengthened main
tether parts are needed.

\begin{figure}
\includegraphics[width=0.5\columnwidth]{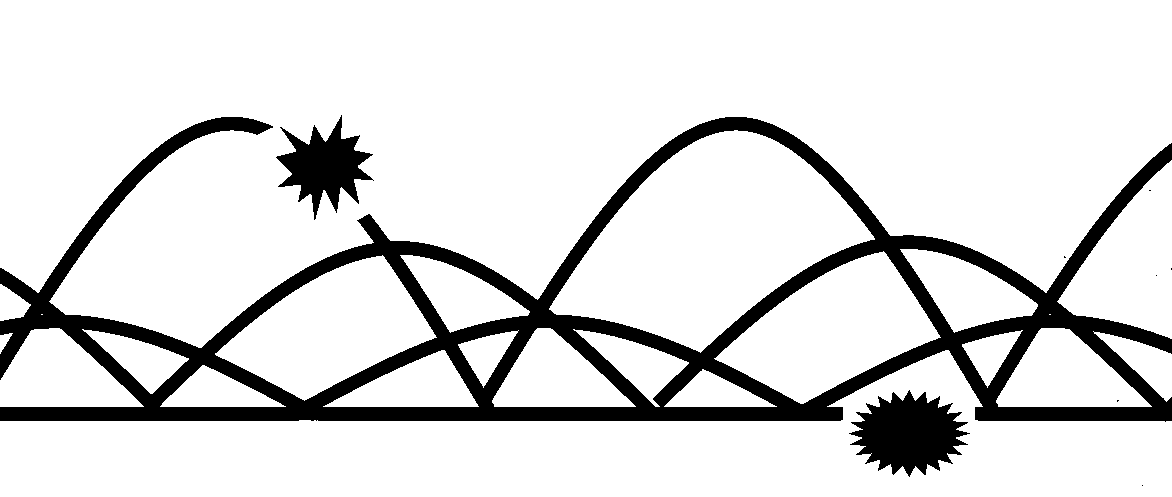}
\caption{
Construction of fourfold micrometeoroid-resistant conducting tether
out of thin aluminium wires. If a micrometeoroid breaks the base wire
or a loop wire, one of the other loop wires takes up the load. The
tether breaks only if all four wires are broken in the same segment.
}
\label{fig:Heytether}
\end{figure}

The E-sail main tether which is currently under development is made of
a 50 $\mu$m aluminium base wire onto which three sequences of loop
wires of 25 $\mu$m thickness are bonded ultrasonically at 2-3 cm
spacing (Fig.~\ref{fig:Heytether} \cite{SeppanenEtAl2011}). The loop wires are added to
increase the survivability of the tether against micrometeoroid
impacts. The base wire is thicker than the loop wires to ease
manufacturing. The pull strength of the tether is the same as the
breaking strength of the ultrasonic 25 $\mu$m on 50 $\mu$m bond which
is about 11 cN. With safety margins, a tether tension of 5 cN can be
used which corresponds to about 1 cN maximum solar wind force from one
tether. According to plasma simulations and theoretical estimates, the
expected E-sail force per tether length at 1 AU and at 20 kV tether
voltage is 0.5 $\mu$N/m \cite{paper6}. The E-sail force scales roughly
linearly with the voltage. For fixed tether voltage it scales as the
square root of the solar wind dynamic pressure so that its dependence
on the solar radial distance $r$ is $1/r$. Notice that the E-sail
force decays slower with $r$ than the photon pressure which scales as
$1/r^2$. The reason for the different scaling is that the virtual
``sail area'' of the E-sail tether is proportional to the solar wind
plasma Debye length which scales as \MARKI{$1/\sqrt{n}$} where $n$ is the solar
wind plasma density. The Debye length therefore scales as proportional
to $r$ which partly cancels the $1/r^2$ scaling of the solar wind
dynamic pressure (the E-sail force per tether length is proportional
to the dynamic pressure and the Debye length, the coefficient of
proportionality depending on the tether voltage and other parameters).

\begin{center}
\begin{table}
\begin{tabular}{lll}
Parameter                    & Symbol    &   Value \\
\hline
Number of tethers            & $N$       & 100 \\
Main tether length           & $R$       & 20 km \\
\MARKI{E-sail force per length at 1 AU}&    & \MARKI{500 nN/m} \\
Spin period \MARKI{at 1 AU}  &           & 60 min \\
Diameter of base wire        & $r_w^{\rm base}$ & 50 $\mu$m \\
Diameter of loop wire        & $r_w^{\rm loop}$ & 25 $\mu$m \\
Main tether mass per length  & $\lambda$ & 11.4 g/km \\
Total mass of main tethers   &           & 22.7 kg \\
Aux.~tether thickness        &           & 7.6 $\mu$m \\
Aux.~tether tape width       &           & 3 cm \\
Aux.~tether punching         &           & 50\% \\
Aux.~tether mass per length  & $\lambda_{\rm aux}$& 163 g/km \\
Total mass of aux.tethers    &           & 20.5 kg \\
\MARKI{Tether tension at 1 AU}&          & \MARKI{0.05 cN} \\
Blade thrust required \MARKI{at 1 AU} &  & 5.4 $\mu$N \\
Blade area required          &           & 2.7 m$^2$ \\
Blade mass (7.6 $\mu$m kapton) &           & 29 g \\
Total mass of Remote Units   &           & 50 kg \\
\hline
\end{tabular}
\caption{
Parameters of E-sail with auxiliary tethers, Remote Units and
photonic blades.
}
\label{table:table1}
\end{table}
\end{center}

Table \ref{table:table1} gives typical parameters for an E-sail having
2000 km total tether length (100$\times$20 km) and producing 1 N
thrust at 1 AU. The minimum blade area has been calculated \MARKI{from
the requirement that it is able to cancel the Coriolis acceleration
(\ref{eq:coriolis1}) in circular orbit. We} assume
90\% reflecting photonic sail material. When the E-sail is inclined to
a typical 45$^{\rm o}$ angle with respect to the solar wind flow, the
photonic thrust which is available to accelerate or decelerate the
spin is about 25\% of the full photonic \MARKI{force per area (8.17
  $\mu$N/m$^2$ at 1 AU)} that would be obtained
by placing \MARKI{the 90\% reflecting} blade perpendicular with respect to the sun direction;
this factor was assumed when computing the required blade area in
Table \ref{table:table1}. \MARKI{The spin period is adjusted so as to keep
  the tether tension five times
  larger than the nominal E-sail force at each distance.
  According to our dynamical simulations using realistic solar
  wind data, such tether tension is enough to keep the tether rig
  stable when auxiliary tethers are used.}.

If the \MARKI{tether} voltage is \MARKI{kept} constant, the E-sail thrust scales as $1/r$ where
$r$ is the solar distance. To keep the ratio of the E-sail force and
the centrifugally produced tether tension constant, the sail spin rate
$\omega$ must changed as $1/\sqrt{r}$. On a quasi-circular orbit the
orbital angular frequency $\Omega$ scales as $r^{-3/2}$ so that the
secular angular acceleration of the sail due to the orbital Coriolis
effect Eq.~(\ref{eq:coriolis2}) scales as $\sim \omega \Omega \sim
1/r^2$. This secular acceleration must be compensated by the photonic
blade so that the blade thrust must scale as $1/r^2$. Luckily this is
the same scaling than that of the solar radiation pressure. Thus we
conclude that if the blade is \MARKI{sized} to compensate the orbital
Coriolis effect at 1 AU, it works at other solar distances as well, at
least under the assumption of nearly circular orbits. Although we do
not prove it rigorously, it seems likely that the circular orbit
represents the worst case for the orbital Coriolis effect so that the
blade dimensioning approach described above should be sufficient for
any E-sail mission, regardless of the orbit.

Notice \MARKI{in Table \ref{table:table1}} that the photonic blade mass (29 g) is an insignificant
fraction of the Remote Unit mass (0.5 kg). We assumed 7.6 $\mu$m
kapton for both auxiliary tethers and photonic blades (uncoated in
case of auxiliary tethers, doubly aluminised in case of the blades).

The Remote Unit mass estimate of 0.5 kg is derived from ongoing
prototyping work in the ESAIL project where the cold gas thruster
based Remote Unit's current dry mass estimate is 0.55 kg. We estimate
that the photonic blade version of the Remote Unit would be slightly
lighter than the one containing butane tank and two cold gas
thrusters, because the mass of the blade itself is not more than
50-100 g (even smaller if advanced thin material is used) and the mass
of its tilting mechanism is also not large. From Table 1, the total
mass of the tether rig (main tethers, auxiliary tethers and Remote
Units with their blades) comes out as 93 kg. Adding to that the mass
of the main tether reels, the high voltage subsystem with the electron
guns and necessary sensors and control electronics on the main
spacecraft would bring the total 1 N E-sail propulsion system mass
into the 120-150 kg range. Reducing the total mass to 100 kg seems not
to be out of question either. Taking into account that a 1 N E-sail
can carry a 1000 kg total mass with reasonably good nominal
acceleration of 1 mm/s$^2$.  For such applications, reducing the
E-sail's own mass by 30 kg (25\%), say, yields only a 9 \% increase in
the payload capability.

The operating principle of the photonic blades described above are
similar to the heliogyro photonic sail blades
\cite{MacNeal1967,Blomqvist2009}. A difference is that in the
traditional heliogyro the blades are tilted mechanically from the main
spacecraft while in our concept they are tilted from the Remote Units.

\section{Leaving out auxiliary tethers: \MARKI{solar fins}}

\begin{figure}
\includegraphics[width=0.65\columnwidth]{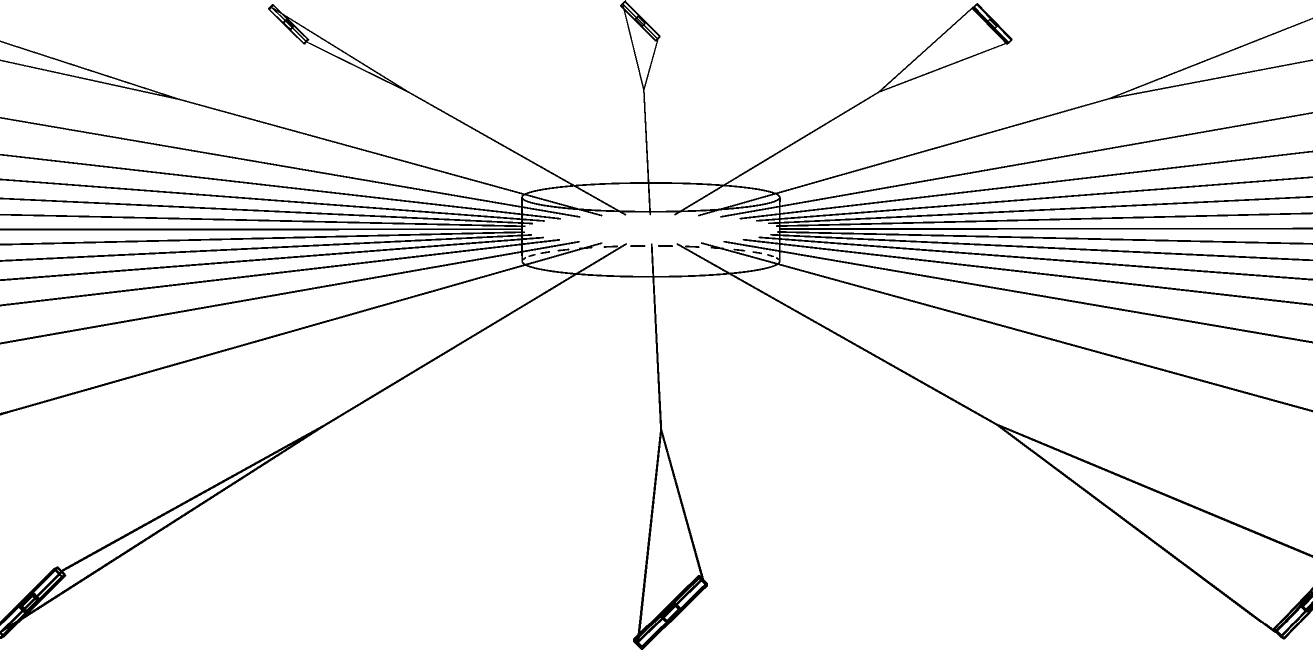}
\caption{
Perspective view of photonically controlled E-sail without auxiliary
tethers. At the end of each main tether there is a triangular solar
\MARKI{fin} whose attitude is controlled by moving a ballast mass along the
bar-like Remote Unit.
}
\label{fig:siipake3D}
\end{figure}

Once the Remote Unit is equipped with a propellantless propulsion
device, it should be possible at least in principle to control its
rotational state by the E-sail and photonic forces alone, without any
help from auxiliary tethers. In other words, in this concept each
tether would be \MARKI{equipped with a solar fin and it would be}
dynamically independent of other tethers
(Fig.~\ref{fig:siipake3D}). \MARKI{This requires that the fin angles
  are continuously adjusted to obtain photonic propulsion not only to
  compensate the Coriolis effect (\ref{eq:coriolis1}), but also to
  ensure that the tethers do not collide with each other although
  solar wind variations and other perturbations introduce small
  differences in their angular speeds.}

\MARKI{In the solar wind concept, the main spacecraft has an imaging
  system for measuring the instantaneous position of each Remote Unit,
  while each Remote Unit has a local sun sensor for measuring its angle of
  rotation about the tether direction. The main spacecraft
  commands the fin angle of each Remote Unit by radio while
  the Remote Unit has a local closed loop controller adjusting the ballast mass
  until the fin's tilting angle is correct.}

The benefits of the \MARKI{solar fin} approach include simplicity and
mass saving (no auxiliary tethers, reels and motors) and the
possibility to scale the system by reducing the number of tethers
without necessarily making them shorter (in designs containing
auxiliary tethers, such combination of parameters would be
uneconomical because it would make the individual auxiliary tethers
long and thus increase the size and mass of their storage reels on the
Remote Units). Also the possibility to test and demonstrate the
dynamical controllability of the tether rig in orbit with a small
number of tethers (minimally only one) would be a behefit of the
independent tether approach.

\begin{figure}
\includegraphics[width=0.75\columnwidth]{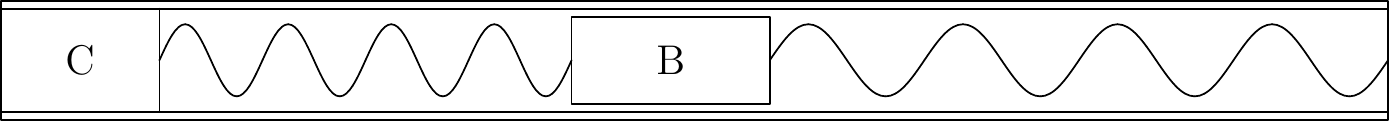}
\caption{
The tube-shaped body of the Remote Unit of the \MARKI{solar fin} E-sail concept
(Fig.~\ref{fig:siipake3D}), with controller ($C$) and battery unit ($B$) which also acts as ballast
mass for changing the centre of mass. The battery unit $B$ is moved by a
linear motor (not shown) and electrically connected to the ends of its tubular
cavity by thin metal springs.
}
\label{fig:RUtube2D}
\end{figure}

Once the Remote Unit does not have auxiliary tethers connected, there
is more freedom to select its shape. In particular, we consider that
the Remote Unit could be a hollow tube around which the \MARKI{solar fin}
is initially wrapped (Fig.~\ref{fig:RUtube2D}). The \MARKI{solar fin} could be
made triangular (Fig.~\ref{fig:siipake3D}) instead of rectangular so
that no other rigid bars are required. The attitude of the \MARKI{fin} could
be controlled by actively changing the centre of mass of the Remote
Unit. The actuation mechanism could be e.g.~a ballast mass that can be
moved along the tube by a linear motor. The Remote Unit would be
powered by solar panels either mounted on its surface or by thin film
solar panels attached on a portion of the photonic \MARKI{fin}. For
robustness, a battery or other energy storage device would be
beneficial to ensure that the unit remains functional in any
orientation with respect to the sun. For mass optimisation the energy
storage device could be part of the ballast mass and the necessarily
electric connections could be made e.g.~by thin cylindrical metallic
springs connecting the ballast mass to the ends of the tube
(Fig.~\ref{fig:RUtube2D}). In this way the only sliding contact would
be the one associated with the linear motor. In desired, the linear
motor could employ magnetic levitation techniques to eliminate sliding
contacts completely. The Remote Unit needs to control its orientation
actively and continuously by adjusting the ballast mass position. The
required control algorithm needs input from local sun sensors as well
as transmitted positional information and commands from the main
spacecraft.

\section{Discussion and outlook}

The idea of using \MARKI{small} photonic \MARKI{sails} for generating the initial E-sail
spin and for modifying the spinrate later during E-sail flight (to
overcome the orbital Coriolis effect or for mission specific reasons)
seems feasible and promising. The required blade size is modest
so that the blade can be made lightweight, without the absolute necessity of
relying on advanced thin photonic sail materials. The propellantless
character of the photonic blades is a benefit which suits
well with the propellantless nature of the E-sail itself.

The solution with auxiliary tethers needs one mechanical actuator per
Remote Unit for turning the photonic blade. This actuator has to
remain functional throughout the E-sail mission. However, failure of a
single Remote Unit (either the whole unit or its tilting mechanism) is
not mission critical because the only requirement is that the total
spinrate control authority provided by the intact Remote Units is
sufficient for the mission.

The \MARKI{solar fin} solution with \MARKI{no} auxiliary tethers
incorporates some way of modifying the centre of mass of the bar-like
Remote Unit such as a sliding mass moved by a linear motor inside a
tubular unit. The latter solution might also be used in a photonic
heliogyro sail, as an alternative to the traditional heliogyro layout
where the \MARKI{solar fins} are rectangular and actuated from the
main spacecraft. In a traditional heliogyro the sum of the widths of
the photonic \MARKI{fins} is limited by the perimeter of the main
spacecraft. By placing the \MARKI{solar fins} on tips of (longer or shorter)
tethers as in Fig.~\ref{fig:siipake3D}, this geometric limitation is
relieved. The length of the tubular and stiff Remote Units define the
width of the \MARKI{solar fins}. This dimension cannot be longer, of course, than
the maximum linear dimension available in the launch vehicle, but the
bar-like Remote Units can be packed at the main spacecraft in more
than one way.

Future engineering and experimental work is needed to study and
develop both concepts further. Especially the robustness and
reliability of the moving parts in both concepts should be analysed in
more detail, and magnetic bearing options should be investigated to
see if sliding tribological contacts could be avoided. \MARKI{For the
  solar fin concept, the question about sufficient fin size to
  overcome risk of tether collisions due to solar wind variations
  should be studied by numerical experiments with a dynamical
  simulator fed by realistic solar wind data.}

\section{Acknowledgement}


The research leading to these results has received funding from the
European Community's Seventh Framework Programme ([FP7/2007-2013])
under grant agreement number 262749. We also acknowledge the Academy
of Finland and the Magnus Ehrnrooth Foundation for financial support.










\end{document}